# OPERATOR FORMALISM FOR OPTICAL NEURAL NETWORK BASED ON THE PARAMETRICAL FOUR-WAVE MIXING PROCESS

L.B. Litinskii, B.V. Kryzhanovsky

Center of Optical Neural Technologies, Scientific Research Institute for System Analisys, Russian Academy of Sciences
44/2 Vavilov Street, 119333 Moscow, e-mail: litin@mail.ru

and

A. Fonarev

Department of Engineering Science and Physics, the College of Staten Island,
CUNY, 2800 Victory Blvd., SI, New York 10314, e-mail: afonarev@gmail.com

**ABSTRACT**

In this paper we develop a formalism allowing us to describe operating of a network based on the parametrical four-wave mixing process that is well-known in nonlinear optics. The recognition power of a network using parametric neurons operating with $q$ different frequencies is considered. It is shown that the storage capacity of such a network is higher compared with the Potts-glass models.

## 1. INTRODUCTION

In the papers [1,2] a network based on the well-known in nonlinear optics parametrical four-wave mixing process (FWM) [3] was examined. Such a network is capable to hold and handle information that is encoded in the form of the phase-frequency modulation. In the network the signals propagate along interconnections in the form of quasi-monochromatic pulses at $q$ different frequencies $\{\omega_k\}^q \equiv \{\omega_1, \omega_2, ..., \omega_q\}$. The model is based on a «parametrical neuron» that is a cubic nonlinear element capable to transform and generate frequencies in the adiabatic parametrical FWM processes [4-6]:

$$\omega_i - \omega_j + \omega_k \to \omega_r .$$

Schematically this model of a neuron can be assumed as a device that is composed of a summator of input signals, a set of $q$ ideal frequency filters $\{\omega_k\}^q$, a block comparing the amplitudes of the signals and $q$ generators of quasi-monochromatic signals $\{\omega_k\}^q$.

The network operates as follows: the input signals are summarized; the summarized signal propagates through $q$ parallel frequency filters; the output signals from the filters are compared with respect to their amplitudes; the signal with the maximal amplitude activates generation of an output signal whose frequency and phase are the same as the frequency and the phase of the initiating signal. For this scheme the condition of the frequencies noncommensurability is of principle:

$$\omega_i - \omega_j + \omega_k \in \{\omega_r\}^q \text{ only if } \omega_j = \omega_i \text{ or } \omega_j = \omega_k .$$

The fulfillment of this condition guarantees a high level of the internal noises suppression.

In this network the signals are characterized by a set of parameters (frequencies), in other words by a vector. Consequently, the state of neurons has to be described with the aid of vector quantities. Analyzing the approach of refs. [1,2] it can be seen that it is close to the well-known Potts model [4,5] as well to another vector models of neural networks [7-12]. Here we present a general formalism for the model [1] described above.

## 2. OPERATOR FORMALISM

We use the Hopfield network formalism when describing the optical neural network based on the processes of electromagnetic waves interaction in nonlinear medium because of the deep resemblance of the Hopfield and FWM Hamiltonians. Really, let us consider the Hopfield network with $p$ stored patterns

$\vec{X}_\mu = (X_{\mu 1}, X_{\mu 2}, \ldots, X_{\mu N})$, $\mu = 1,2,..p$. Suppose that the state of the network is described by *N*-dimensional vector $\vec{X} = (X_1, X_2, \ldots, X_N)$. Then the Hamiltonian of the network is

$$H = -\sum_{i,j}^{N} X_i T_{ij} X_j^+ \quad (1)$$

where the interconnections are defined as

$$T_{ij} = \sum_{\mu=1}^{p} X_{\mu i}^+ X_{\mu j}, \quad i \neq j; \quad T_{ii} = 0 \quad (2)$$

For the *i*-th neuron the local field is

$$h_i = \sum_{j=1}^{N} T_{ij} X_j^+ \quad (3)$$

where the superscript "+" denotes the Hermitian conjugate. In what follows it will become clear why the Hermitian conjugate is necessary.

The analogy with FWM process becomes clear, if we treat the quantities $X_i$ and $X_{\mu i}$ as amplitudes of electromagnetic waves propagating along interconnections. Here we interpret the interconnections as a nonlinear medium where a residual perturbation $T_{ij}^{(\mu)} = X_{\mu i}^+ X_{\mu j}$ is induced by the waves $X_{\mu i}$ and $X_{\mu j}$. In this approach we interpret the local field as a medium polarization induced under the influence of the wave $X_j$ propagating from the *j*-th neuron to the *i*-th one. Thus we can construct a neural network where the neurons exchange the signals in the form of quasi-monochromatic pulses at *q* different frequencies $\{\omega_k\}^q$. To describe such an optical neural network we need to treat a neuron as an object which has one ground state $|0\rangle$ and *q* excited states $|k\rangle$, $k = 1,\ldots,q$. An analogy is an atom with one ground and *q* excited levels. The state $|k\rangle$ is excited by the wave with the corresponding frequency $\omega_k$ only. The transition $|k\rangle \to |0\rangle$ is accompanied by emitting the wave of frequency $\omega_k$.

A detailed description of such optical neural network processing is given in refs.[1,2]. Herein we will show only that a formal description of this network can be done in the framework of the Hopfield model if we introduce the creation and annihilation operators of neuron states. The properties of these operators are given by the commutation relations:

$$c_k c_l^+ - c_l^+ c_k = \delta_{kl}.$$

The effect of the field on the neuron is described by expressions:

$$c_k |l\rangle = \delta_{kl} |0\rangle, \quad c_k^+ |0\rangle = |k\rangle,$$

$$c_k |0\rangle = 0, \quad c_k^+ |l\rangle = 0.$$

The rule of the multiplication of the vectors is $\langle k|l\rangle = \delta_{kl}$. In Eqs.(1-3) in place of the amplitudes $X_i$ and $X_{\mu i}$ let us substitute the operators $X_i = x_i c_k^{(i)}$ and $X_{\mu i} = x_{\mu i} c_k^{(\mu i)}$, where $x_i, x_{\mu i} = \pm 1$ are the amplitudes of the waves and $c_k^{(i)}, c_k^{(\mu i)} \in \{c_k\}^q$. As a result of this substitution we obtain the Hamiltonian operator and the local field operator that are sufficient for the description of the network processing. For example, to describe the dynamics of the network we have to average the Hamiltonian operator with respect to the ground state of the system $|0\rangle$. Then for the obtained quantities we use the standard analysis in the framework of the Hopfield model. The state of the *i*-th neuron is obtained by the action of the local field operator $h_i$ on the state $|0\rangle$:

$$h_i |0\rangle = \sum_{k=1}^{q} A_k^{(i)} |k\rangle \quad (4)$$

where

$$A_k^{(i)} = \sum_{j \neq i}^{N} \sum_{\mu=1}^{p} x_{\mu i}^* x_{\mu j}^* x_j \langle k | k_{\mu i} \rangle \langle k_{\mu j} | k_j \rangle \quad (5)$$

We see that as a result of local field action the *i*-th neuron turns out to be in a mixed excited state. To determine the type of the signal this neuron emits when relaxing into the ground state, in ref.[1] the rule " winner - take - all" was used: In the series (4) the state with the maximal amplitude is determined; the frequency of this state and the phase (the sign) that coincide with the phase of the amplitude of this state are assigned to the emitted signal. It is easy to see that this rule ensures the system energy decreasing. The further analysis of the capability of the network to recognize the stored patterns is the same as for the standard network of the Hopfield type. In the next section we present a formalized form of such analysis.

### 3. VECTOR FORMALISM

Let us examine a network consisting of $N$ neurons connected with each other. In order to describe the states of neurons we use the set of basic vectors $\vec{e}_k$ in the space $R^q$, where $q \geq 1$. The state of the $i$-th neuron is described by a vector $\vec{x}_i = x_i \vec{e}_k$, where $x_i = \pm 1$, $\vec{e}_k \in R^q$, $k = 1,2,..q;\ i = 1,2,...,N$. Then, the state of the network as a whole $\vec{X}$ is determined by a set of $N$ $q$-dimensional vectors $\vec{x}_i$: $\vec{X} = (\vec{x}_1, \vec{x}_2,..., \vec{x}_N)$. Since in this model neurons are vectors, the local field $\vec{h}_i$ affecting the $i$-th neuron is the vector too. By analogy with the standard Hopfield model, Eqs.(1)-(3), we write

$$\vec{h}_i = \sum_{j=1}^{N} \mathbf{T}_{ij} \vec{x}_j \qquad (6)$$

The $(q \times q)$-matrix $\mathbf{T}_{ij}$ describes the interconnection between the $i$-th and the $i$-th neurons. The elements of this matrix can be calculated with the aid of the generalized Hebb rule:

$$T_{ij}^{(kl)} = \sum_{\mu=1}^{p} (\vec{e}_k \vec{x}_{\mu i})(\vec{x}_{\mu j} \vec{e}_l), \quad T_{ii}^{kl} \equiv 0 \qquad (7)$$

where $p$ stored patterns are $\vec{X}_\mu = (\vec{x}_{\mu 1}, \vec{x}_{\mu 2},..., \vec{x}_{\mu N})$, $\mu = 1,2,...,p$. With regard to Eqs.(6) and (7) the Hamiltonian of the system takes the form:

$$H = -\tfrac{1}{2} \sum_{i,j}^{N} \sum_{\mu=1}^{p} (\vec{x}_i \vec{x}_{\mu i})(\vec{x}_{\mu j} \vec{x}_j)$$

Let us define the dynamics of our $q$-dimensional neurons. Let $\vec{X} = \vec{X}(t)$ be the state of the system at the time $t$. We rewrite the local field (6) affecting the $i$-th neuron at the time $t$ in the form convenient for the further analysis:

$$\vec{h}_i(t) = \sum_{k=1}^{q} A_k^{(i)} \vec{e}_k \qquad (8)$$

where

$$A_k^{(i)} = \sum_{j(\neq i)}^{N} \sum_{\mu=1}^{p} (\vec{e}_k \vec{x}_{\mu i})(\vec{x}_{\mu j} \vec{x}_j) \qquad (9)$$

and $\vec{x}_j = \vec{x}_j(t)$. We use the subscript *max* to denote amplitude $A_k^{(i)}$ in the series (8) that is maximal in modulus. Then the neuron dynamics is as follows: due to the action of the local field the $i$-th neuron at the time $t+1$ is oriented along the basic vector $\vec{e}_{max}$, if $A_{max}^{(i)}(t) \geq 0$, and it is oriented in the opposite direction if $A_{max}^{(i)}(t) < 0$, that is

$$\vec{x}_i(t+1) = \mathrm{sgn}[A_{max}^{(i)}(t)] \vec{e}_{max} \qquad (10)$$

The dynamics of the system consists of consequent variations of the values of neurons according to the rule (10). This dynamics corresponds to decreasing of the energy of the system during its processing (a "spin" is oriented along a direction mostly close to the external field).

In conclusion we would like to note that the vector model describes completely the behavior of the operator model of neural network discussed in the previous section. The expression (10) is identical to the "winner – take – all"-rule of section 2, and Eqs.(4) and (5) can be transformed in Eqs.(8) and (9) with the replacement of the state vectors $|k\rangle$ by the basic vectors $\vec{e}_k$. Consequently, the following analysis of the storage capacity of the network is valid for both models.

### 4. STORAGE CAPACITY OF NETWORK

It can be shown that if the number of the stored patterns is equal to one or two, $p \leq 2$, the fixed points of the network are coinside with the stored patterns only. Thus, if $p \leq 2$ our model behaves as the standard Hopfield model.

Let us estimate the storage capacity of the network for an arbitrary value of $p$ in the limit $N \to \infty$. Suppose that the network starts from a distorted pattern

$$\vec{X} = (a_1 b_1 \vec{x}_{m1}, a_2 b_2 \vec{x}_{m2}, ..., a_N b_N \vec{x}_{mN}) \qquad (11)$$

Here $\vec{X}_m = (\vec{x}_{m1}, \vec{x}_{m2}, ..., \vec{x}_{mN}) \in \{\vec{X}_\mu\}^q$ is one of the stored patterns, $\{a_i\}^N$ and $\{b_i\}^N$ define a multiplicative noise: $a_i$ is a random value that is equal to -1 or +1 with the probabilities $a$ and $1-a$ respectively; $b$ is the probability that the operator $b_i$ changes the state of the vector $\vec{x}_{mi}$, and $1-b$ is the probability that vector $\vec{x}_{mi}$ remains unchanged.

Let us examine to what extent the neural network recognize the vector $\vec{X}_m$ correctly. At first, let us note that when the patterns $\vec{X}_\mu$ are uncorrelated, the quantities of the types $\xi = a_j (\vec{x}_{mj} b_j \vec{x}_{mj})$ and $\eta = a_j (\vec{e}_k \vec{x}_{\mu i})(\vec{x}_{\mu j} b_j \vec{x}_{mj})$ with $\mu \neq m$ can be considered as independent random variables that are described by the probability distributions

$$\xi_r = \begin{cases} +1, & (1-a)(1-b) \\ 0, & b \\ -1 & (1-b)a \end{cases}$$

$$\eta_r = \begin{cases} +1, & 1/2q^2 \\ 0, & 1-1/q^2 \\ -1 & 1/2q^2 \end{cases}$$

Then substituting Eq.(11) in Eq.(9) we obtain

$$A_k^{(i)} = \vec{e}_k \vec{x}_{mi} \sum_{r=1}^{N-1} \xi_r + \sum_{r=1}^{L} \eta_r, \quad \text{when } \vec{e}_k \vec{x}_{mi} \neq 0 \qquad (12)$$

$$A_k^{(i)} = \sum_{r=1}^{L} \eta_r, \qquad \text{when } \vec{e}_k \vec{x}_{mi} = 0 \qquad (13)$$

where $L \equiv (N-1)(p-1)$. According to the rule (10) the $i$-th neuron finds itself in the state $\vec{x}_{mi}$ when two conditions are fulfilled: *i)* the amplitude (12) in modulus is greater than all the other amplitudes (13); *ii)* the sign of the amplitude (12) is the same as the sign of $x_{mi}$. Otherwise there is an error of the vector $\vec{x}_{mi}$ recognition. The probability of this error is

$$\Pr_i = \Pr\left\{ \sum_{r=1}^{N-1} \xi_r \leq \sum_{r=1}^{L} \eta_r \right\}$$

To estimate the value of $\Pr_i$ we use the well-known Chernov – Chebyshev method [15]. As a result we obtain the expression for the probability of the error of the vector $\vec{X}_m$ recognition:

$$\Pr = N \exp\left[ -\frac{Nq^2}{2p}(1-2a)^2 (1-b)^2 \right] \qquad (14)$$

When $N$ increases, this probability tends to zero, if $p$ as function of $N$ increases slower than

$$\bar{p} = \frac{N}{2\ln N} q^2 (1-2a)^2 (1-b)^2 \qquad (15)$$

This allows us to use (15) as an asymptotically possible value of the storage capacity of our neural network.

## 5. DISCUSSION

As it follows from (15), the network is more sensitive to an error in the sign than in the number of the state. When $q=1$, Eqs.(14)-(15) transform into well-known results for the standard Hopfield model. When $q$ increases the noise immunity of the considered associated memory increases noticeably. In the same time the storage capacity of the network increases proportionally to $q^2$. Moreover, in contrast to the Hopfield model the number of the patterns $p$ can many times exceed the number of neurons. The aforementioned properties of our model are due to the complex structure of neurons from which the network is composed.

Let us compare our model with the Potts-glass model. When $q=2$, the last also transforms into the Hopfield model. According to the estimates of ref.[7] the storage capacity of the Potts neural network is $\alpha_c(q) = \frac{1}{2} q(q-1)\alpha_0$, where $\alpha_0 \approx 0.138$. In our model for $q \geq 1$ we have $\alpha_c(q) = q^2 \alpha_0$. We see that if $q$ is sufficiently large in our model the storage capacity is twice as big as the storage capacity of the models [7-14]. We think that the gain is achieved because of the frequencies noncommensurability principle (see Introduction), which was not used in ref.[7-14]. The comparison allows us to conclude that realization of the optical neural network based on the parametrical four-wave mixing process can be more efficient than the neural network based on the Potts-glass model.

Summarizing the data, we can conclude the following: 1) introduction of frequency characteristics for the components of the processed images leads to a significant increase in the volume of neural memory and reduce recognition errors. 2) the number of interconnections in the resulting of «channel multiplexing» can be reduced in $q^2$ time, without reducing the sizeof the memory and without increasing the error detection, i.e. to a certain extent to solve the problem of $N^2$. 3) the memory size can be varied without changing the dimensions of the processed vectors, i.e. by fixing the value of N. 4) frequencies can encode a variety of feature such as color pixel images, etc. Of course, the complexity of the neuron results in an increase in the number of local connections within itself. However, more importantly, this decreases the number of long-distance interconnections to other neurons.

In conclusion, we would like to note that the presented approach allows us to describe not only the optical neural networks of the parametrical type, but also neural networks in which information is encoded in the form of phase delays of pulses in interconnections. It is much more easy to realize such a network in form of a device.